\documentclass[eprint,amsmath,amssymb,aps, superscriptaddress]{revtex4-2}
\usepackage{mathtools,bm}
\usepackage{graphicx}
\usepackage{hyperref}
\hypersetup{colorlinks=True,
citecolor=red,
linkcolor=blue, 
urlcolor=black}
\usepackage[dvipsnames]{xcolor}
\newcommand{\deriv}{\mathrm{d}}

\begin{document}

\preprint{APS/123-QED}

\title{Obstacle-induced lateral dispersion and nontrivial trapping of flexible fibers settling in a viscous fluid}

\author{Ursy Makanga}
\author{Mohammadreza Sepahi}
\author{Camille Duprat}
\email{camille.duprat@ladhyx.polytechnique.fr}
\author{Blaise Delmotte}
\email{blaise.delmotte@ladhyx.polytechnique.fr}

\affiliation{LadHyX, CNRS, Ecole Polytechnique, Institut Polytechnique de Paris, 91128, Palaiseau, France.}

\date{\today}

\begin{abstract}
The motion of flexible fibers through structured fluidic environments is ubiquitous in nature and industrial applications. Most often, their dynamics results from the complex interplay between internal elastic stresses, contact forces and hydrodynamic interactions with the walls and obstacles.  By means of numerical simulations, experiments and analytical predictions, we investigate the dynamics of flexible fibers settling in a viscous fluid embedded with obstacles of arbitrary shapes.  We identify and characterize two types of events: trapping and gliding, for which we detail the mechanisms at play. We observe nontrivial trapping conformations on sharp obstacles that result from a subtle balance between elasticity, gravity and friction. In the gliding case, a flexible fiber reorients and drifts sideways after sliding along the obstacle. The subsequent lateral displacement is large  compared to the fiber length and strongly depends on its mechanical and geometrical properties. We show how these effects can be leveraged to propose a new strategy to sort particles based on their size and/or elasticity. This approach has the major advantage of being simple to implement and fully passive, since no energy is needed.
\end{abstract}

\maketitle

\section{Introduction}
The transport and trapping of fibers through complex environments, such as porous media, occurs in a variety of systems. Small fibers, e.g. micro-plastics fibers, may propagate in soil and cause pollution of groundwater \cite{re_shedding_2019,engdahl_simulating_2018}. When flowed through small cracks in natural rocks, such flexible fibers may buckle, leading eventually to clogging and thus closing of small paths \cite{dangelo_single_2010}, which is for example used to prevent proppant flowback in petroleum engineering \cite{corsano_fiberproppant_nodate}. Similar clogging may happen in the vascular system, where biofilm streamers may form in irregular channels, detach, transport and ultimately remain attached and clog small vessels or structures such as stents \cite{rusconi_laminar_2010,drescher_biofilm_2013}. The characterization of the interaction of flexible fibers with obstacles can further be used to design chromatographic devices to separate DNA filaments by size by transporting them through periodic arrays of posts in microfluidic devices \cite{chou1999sorting}. Other industrial processes rely on the trapping of the fibers on the obstacles, e.g. in papermaking where the fibers accumulates on the wires of a fabric through which a suspension is drained in order to form the paper sheet \cite{vakil_flexible_2011}. In all these situations, the velocity, deformation or trapping of the fiber is determined by a complex interplay between elasticity, viscosity, and interactions with the obstacles.

When a rigid fiber is freely transported in a viscous flow, the trajectory of its center of mass globally follows the streamlines. The fiber may further interact hydrodynamically with the bounding walls of the channel, causing rotations and reorientations \cite{nagel_oscillations_2018, cappello_transport_2019,du_roure_dynamics_2019}. When freely transported fibers encounter obstacles, they will thus glide around them \cite{lopez_deformation_2015,chakrabarti_trapping_2020}. When the fiber is flexible, it may deform in response to viscous forces \cite{du_roure_dynamics_2019}. These deformations help the fiber escape and migrate through the flow streamlines, e.g. through buckling \cite{wandersman2010buckled,quennouz_transport_2015}, bending or coiling \cite{xue_shear-induced_2022}. In the presence of obstacles, flexible fibers may thus deform, stretch, buckle, vault and tumble due to the flow generated by the obstacles, which will affect their trajectory as well as their transport time \cite{sabrio_main_2022,kawale_polymer_2017,chakrabarti_trapping_2020}; indeed, these dynamics increase the path taken by the fiber, and may further lead to prolonged trapping periods or fibers remaining trapped on the obstacles \cite{vakil_flexible_2011}. The presence of obstacles thus affects the long-time transport properties of fibers, such as  dispersion, and since the collision times and the associated transport velocity are size-dependent, this can serve as a base for sorting devices. However, in these cases the overall lateral displacement remains small, as freely transported fibers globally tend to align with the flow. 

Conversely, when moving thanks to external forces such as gravity or in a centrifuge, without an external driving flow, the trajectories of a fiber strongly depends on its orientation with respect to the direction of the force. In particular, a rigid fiber settling in a quiescent fluid oriented at a angle neither perpendicularly nor aligned with the force will drift as it settles down, leading to large lateral displacements. 
During settling, flexible fibers experience deformations and reorientations that strongly affect their transport \cite{saggiorato2015conformations,marchetti_deformation_2018,cunha_settling_2022}. Indeed, a flexible fiber bends due to its own hydrodynamic disturbances; this deformation leads to a torque that re-orients the fiber perpendicularly to gravity, thus reducing its lateral drift \cite{xu_deformation_1994,li_sedimentation_2013}, which also affects their collective dynamics \cite{manikantan_instability_2014,schoeller_methods_2021}. In this situation, the presence of an obstacle, by modifying the deformation and orientation of the fiber, will thus directly affect its trajectory. Here, we consider the prototypal case of a single flexible fiber settling in a quiescent fluid embedded with fixed obstacles of various shapes. We show that the fiber interacts with the obstacle both hydrodynamically and through friction. The presence of the obstacle induces a reorientation with an angle that depends on the fiber characteristics, in particular its flexibility, and the obstacle shape. The fiber then reorients to reach its equilibrium shape after having travelled a finite distance; the magnitude of this lateral displacement is large (typically several fiber lengths) and depends on the fiber characteristics, an effect we leverage on to propose a sensitive sorting solution.  Furthermore, we examine the conditions under which a fiber may remain trapped on the obstacle, and identify non-trivial trapping events, providing design clues for optimal sorting/filtration.

\section{Problem description and relevant parameters}
\begin{figure}[h]
   \centering
    \includegraphics[width=\textwidth]{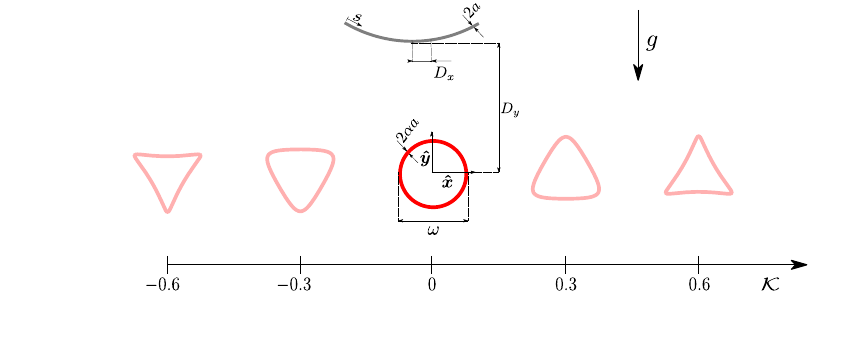}
    \caption{Schematic of the problem considered: a fiber sediments with its equilibrium shape towards an obstacle of cross-section controlled by the parameter $\mathcal{K}$ ranging between  $\mathcal{K} = -0.6$ to $\mathcal{K} = 0.6$.}
    \label{fig: problem_description}
\end{figure}

We study the dynamics of a flexible fiber settling in a viscous fluid, with viscosity $\eta$, embedded with a rigid obstacle, as shown in Fig.~\ref{fig: problem_description}. 
The flexible fiber is  an elastic rod of length $L$ and circular cross-section of radius $a$, with centerline characterized by its arc length $s \in [0, L]$. The fiber is settling at a velocity $U$ such that the Reynolds number $Re=\rho U L/\eta$ is always small and viscous effects dominate the hydrodynamics. The obstacle is defined as a rigid tube of width $\omega$ and depth $D_O$. 
The cross-section of the tube is generated by an area-preserving conformal mapping of the unit circle \cite{avron_optimal_2004, alonso-matilla_transport_2019} (see Appendix \ref{appendix:conformal_mapping}).
The  obstacle shape and symmetry are controlled by a geometric parameter  $\mathcal{K}$: $\mathcal{K}=0$ corresponds to a circular cross-section while $\mathcal{K} = \pm 0.6$ is a triangle with negative curvature pointing upward (downward respectively).
The fiber starts at equilibrium with its midpoint  initially positioned at a given horizontal and vertical distance, $D_x$ and $D_y$,  from the obstacle center of mass. 

In the absence of obstacles, the equilibrium shape of a fiber subject to gravity in a viscous fluid results from the balance between viscous and internal elastic stresses, which is quantified by the so-called elasto-gravitational number $Be=F^G L^2/EI = WL^3/EI$, where $F^G=WL$ is the gravity force, $W$ the weight per unit length of the fiber,  $E$ its Young's modulus  and $I=\pi a^4/4$ its second moment of inertia.
In the rigid case, $Be \ll 1$, the fiber keeps its initial shape, while in the flexible regime, $Be \gg 1$, it bends due to its own hydrodynamic disturbances. The fiber will thus adopt a more or less pronounced ``U" shape, depending on $Be$, oriented perpendicularly with the direction of gravity, independently of its initial configuration, as sketched in Fig.~\ref{fig: problem_description}.
As we will show below, the presence of an obstacle destabilizes this equilibrium shape and affects its trajectory. These changes depend on two additional parameters: the relative length of the fiber with respect to the obstacle width $\xi = L/\omega$ and the obstacle shape $\mathcal{K}$. 

\section{Experimental and numerical methods}\label{sec: methods}

\subsection{Experimental methods}
Experiments are carried out using a slightly modified version of the  experimental setup and protocol described in \cite{marchetti_deformation_2018}.  We fabricate elastic fibers of controlled properties by molding liquid poly-vinyl siloxane (PVS 8, Elite double Zhermack) mixed with iron powder (from $10$ to $20\%$) into a capillary tube of radius $a=510 \;\text{$\mu$m}$; after degassing and cross-linking at room temperature, the fibers are extracted from the tubes. We work with two solutions: $10\%$ Fe, for which $\rho_f=1107 \pm 40$ kg/m$^3$ and $E=141\pm 50\; \text{kPa}$, and $20\%$ Fe for which $\rho_f=1143 \pm 13$ kg/m$^3$ and $E=180\pm 75\; \text{kPa}$. The fibers are then released in a large tank which has a rectangular cross-section ($L_1 = 60\;\text{cm}$, $L_2= 40\;\text{cm}$) and height $H = 80 \;\text{cm}$, filled with silicone oil ($\rho=970$ kg/m$^3$, $\eta=0.97\;\text{Pa.s}$). A 3D-printed obstacle of width 
$1\;\text{cm}$ and spanning the entire depth of the tank is held against the walls in the center of the tank (Fig. \ref{fig: exp_and_num_setup}$(a)$). Fibers are initially held in the upper part of the tank by tweezers in a shape close to their equilibrium configuration. It ensures that  equilibrium is reach before interacting with the obstacles. Fibers are released by slightly opening the tweezers simultaneously to avoid large flow disturbances. We track the shape and position of the fiber using a high-resolution DSLR camera with a wide $20\;\text{mm}$ lens. Typically, the fiber settling velocity is $0.5$ mm/s, corresponding to Reynolds numbers $Re\sim 10^{-2}$, and images are taken every 10s. 

\subsection{Numerical simulations}
\label{sec:numerical_method}
\begin{figure}
   \centering
    \includegraphics[width=\textwidth]{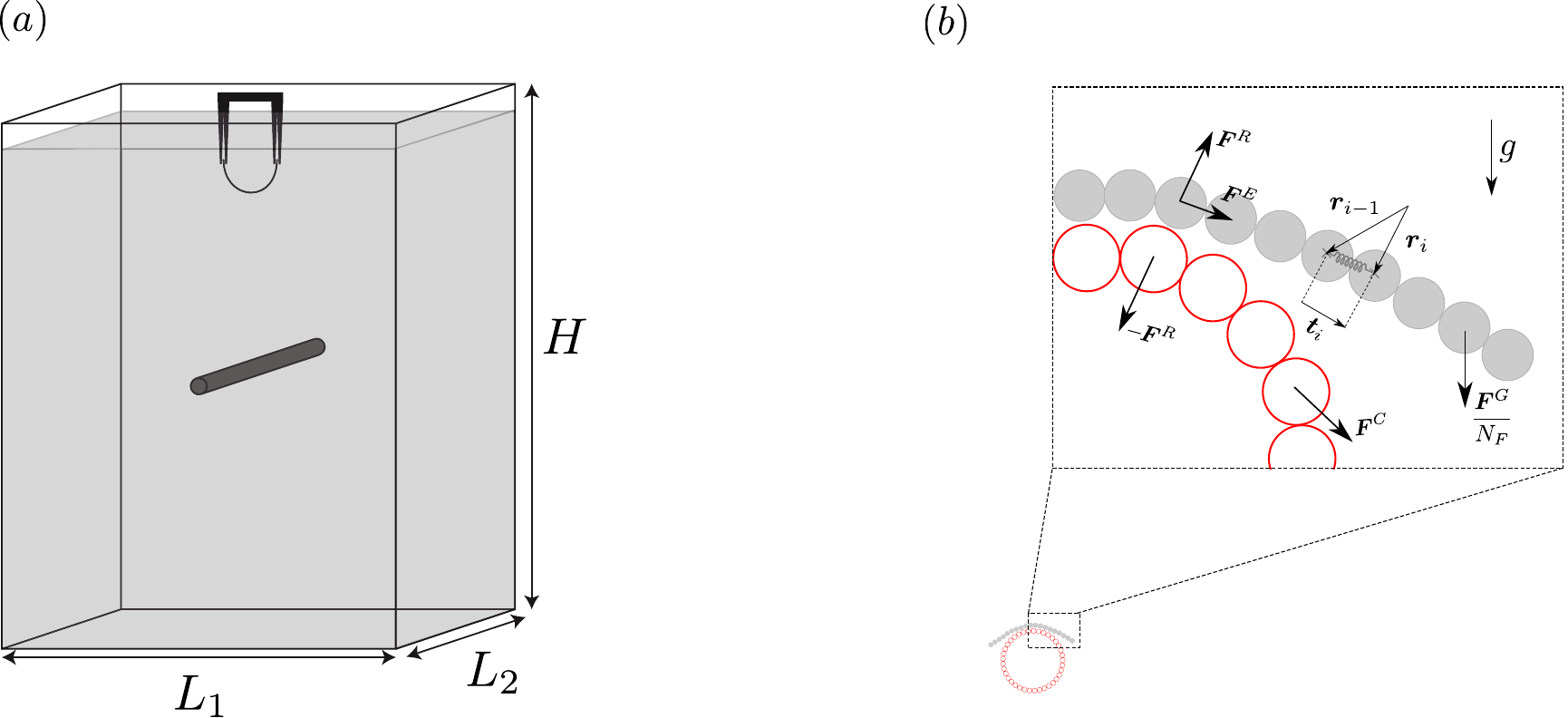}
    \caption{$(a)$ Experimental setup, showing a fiber initially held with tweezers under the free-surface of the tank, with an obstacle of circular cross-section held at the center of the tank. $(b)$ Numerical model, showing the fiber (gray), represented by an inextensible chain of $N_F$ spherical beads linked elastically by Hookean springs, and an obstacle slice (red), discretized by a serie of $N_O$ beads along its contour. Both fiber and obstacle beads are subject to external and internal forces: $\bm{F}^R$ is the repulsive contact barrier force between the fiber and obstacle beads, $\bm{F}^C$ is the constraint force applied to obstacle beads in order to prescribe their motion, $\bm{F}^E$  and $\bm{F}^G/N_F$ are respectively the internal elastic and gravitational forces experienced by the fiber beads. $\bm{t}_i=\bm{r}_i - \bm{r}_{i-1}$ is tangent vector to the fiber centerline at the position of bead $i-1$.} 
    \label{fig: exp_and_num_setup}
\end{figure}

Our numerical method relies on the bead model to solve the elasto-hydrodynamic couplings and contact interactions between fibers and obstacles.
The fiber is modelled as a chain of $N_F$ spherical beads of radius $a$ connected by Hookean springs (see Fig.\ \ref{fig: exp_and_num_setup}$(b)$).
The fiber beads are subject to a gravitational force $\bm{F}^G/N_F$, where $F^G=\|\bm{F}^G\|$ is the weight of the whole fiber.
Mechanical interactions between the fiber beads are governed by elastic forces, $\bm{F}^E$, that derive from the stretching and bending free energies \cite{marchetti_deformation_2018}. Since we are considering plane deformations, twisting of the fiber is neglected. The obstacle surface is discretized with $N_O$ beads of radius $a_O$ stacked in slices. Each slice is discretized with a uniform distribution of beads in contact along its contour $\mathcal{C}$.  
The total number of beads in the system is $N=N_F+N_O$. Since the obstacle is not moving, the obstacle beads must have zero velocity. This kinematic constraint is enforced with a set of constraint forces $\bm{F}^{C}$ applied on the obstacle beads only. The contact forces between the obstacle and fiber beads, $\bm{F}^R$, are pairwise, short-ranged and repulsive. The repulsion between bead $i$ and $j$ is given by  \cite{dance_collision_2004}
\begin{equation}
    \bm{F}_{ij}^{R}=
   \begin{dcases}
     -\frac{F_{R}}{(a_i + a_j)}\left[ \frac{R_{c}^{2} - r_{ij}^{2}}{R_{c}^{2} - (a_i + a_j)^{2}}\right]^{4}\bm{r}_{ij}, & \text{for } r_{ij} < R_{c}\\
      \bm{0}, & \text{otherwise}
\end{dcases}
   \label{eq: steric_def_forces}
\end{equation}
\par
\noindent
where  $\bm{r}_{ij} = \bm{r}_{j} - \bm{r}_{i}$ is the vector between the bead centers and $r_{ij} = |\bm{r}_{ij}|$ . The repulsion strength is chosen to prevent bead overlaps, $F_{R} = 4 F^G$, and $R_{c} = 1.1 (a_i + a_j)$ sets the cutoff distance over which the force acts, here $10\%$ of the contact distance between a pair of beads.
Owing the spherical shape of the beads, and to the discrete nature of the bead model, the repulsive force \eqref{eq: steric_def_forces} can exert tangential efforts along  the fiber centerline. 
The tangential component of $\bm{F}^{R}$, denoted $\bm{F}_{\tau}^{R}$ hereinafter, therefore acts as a friction force between the fiber and obstacle surface. For a given obstacle shape and fiber conformation, the relative magnitude of this friction force is controlled by the relative size of the obstacle beads with respect to the fiber beads, $\alpha =a_O/a$. The maximum penetration of the obstacle bead in the void between fiber beads is given by $\delta/a = 1+\alpha-\sqrt{\alpha(2+\alpha)}$, which is an upper bound of the  effective roughness between the two surfaces. This maximum value is reached when the local radius of curvature of the obstacle is equal to $a_O$, i.e.\ when the obstacle has cusps and the beads are touching. In the following we choose $\alpha = 0.61-2$, which sets the maximum interpenetration to $\delta/a = 0.17 - 0.35$.\\
Because the Reynolds number associated with the fiber motion is relatively small ($Re \ll 1$), the equation of motion of the beads, in the absence of background flow, is given by the linear mobility relation between forces and velocities
\begin{equation}
 \frac{d\bm{R}}{dt} \equiv \bm{U} = \bm{\mathcal{M}} \cdot \left(\bm{F}^{C} + \bm{F}^G/N_F + \bm{F}^E + \bm{F}^R \right)
\label{eq: mob_problem},
\end{equation}
where $\bm{R} = [\bm{r}_1,\cdots, \bm{r}_{N_F}, \bm{r}_{N_F+1}, \cdots,  \bm{r}_{N}]$ and $\bm{U} = [\bm{U}_1,\cdots, \bm{U}_{N_F}, \bm{U}_{N_F+1}, \cdots,  \bm{U}_{N}]$  are $3N$ vectors collecting the bead positions and translational velocities. $\bm{\mathcal{M}}$ is the $3N\times 3N$ bead mobility matrix, which contains their hydrodynamic interactions (HI) in a viscous fluid. In this work, we use the Rotne-Prager-Yamakawa (RPY) matrix which approximates pairwise hydrodynamic interactions between beads of different radii in an infinite fluid domain \cite{zuk_rotneprageryamakawa_2014}. Here the choice of an unbounded geometry is justified by the fact that the tank dimensions are approximately one order of magnitude larger than the fiber length in experiments, and also because the fiber relaxes to equilibrium long before reaching the bottom of the tank.  However, extensions of the RPY mobility matrix to other boundary conditions are available in the literature, such as no-slip boundaries \cite{wajnryb_generalization_2013, swan_simulation_2007}, confined domains \cite{swan_particle_2010} and periodic domains \cite{mizerski_rotne-prager-yamakawa_2014, fiore_rapid_2017}.
After computing $\bm{U}$ we then integrate the bead positions in the equation of motion  \eqref{eq: mob_problem} using an implicit time integrator based on Backward Differentiation Formula (BDF) with adaptative time-stepping \cite{brown_vode_1989}.
In our simulations, all the matrix vector products  (e.g.\ $\bm{\mathcal{M}} \cdot \bm{F}$) are computed with Graphic Processing Units (GPU's), which allows to handle large number of beads with a low computational cost and to achieve linear scaling up to $N\approx 10^4$ \cite{usabiaga_hydrodynamics_2016}.

\section{Results and Discussion}

\begin{figure}[ht]
   \centering
    \includegraphics[width=\textwidth]{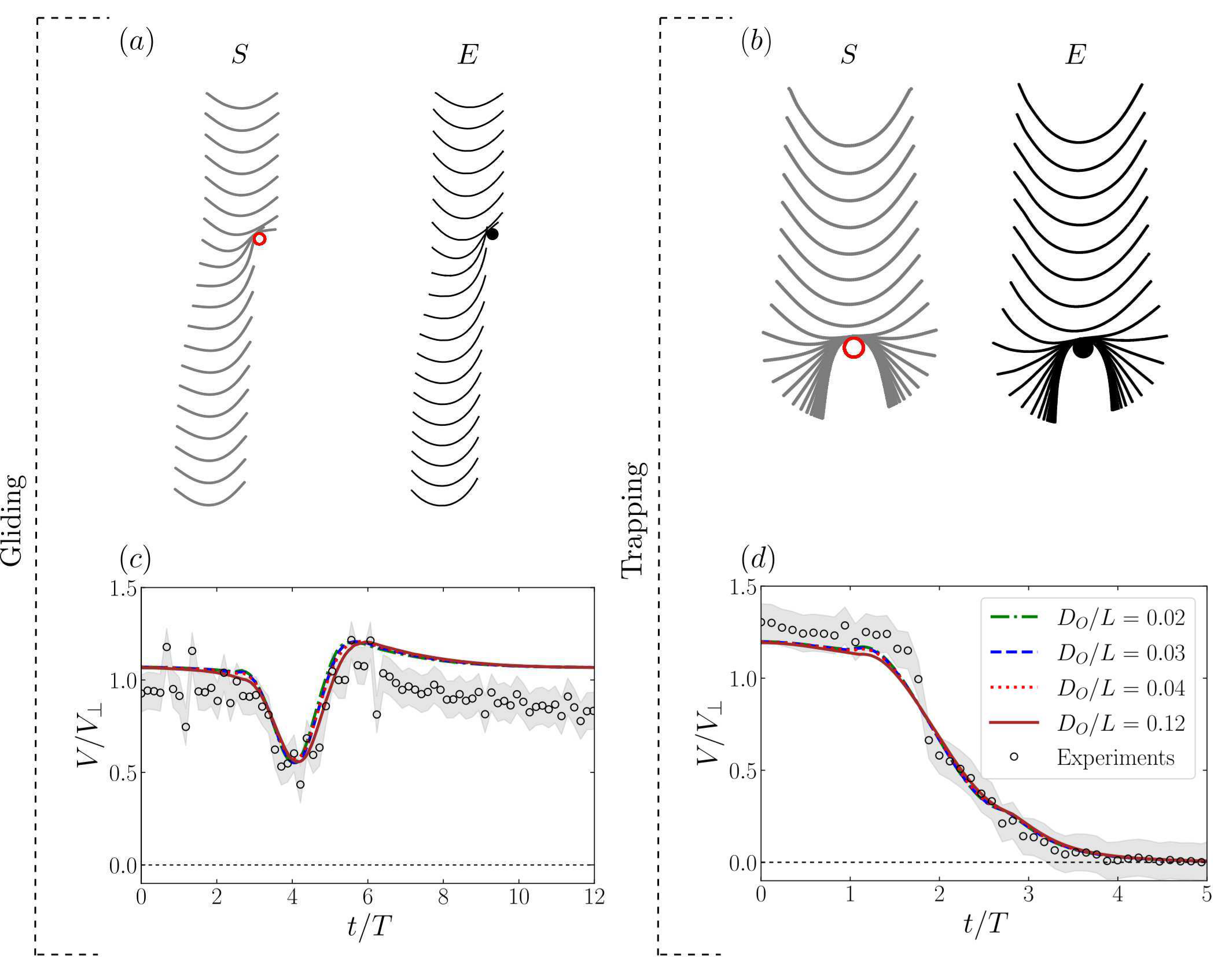}
    \caption{$(a-b)$ Numerical (S) and experimental (E) chronophotographies of a flexible fiber settling against an obstacle in a viscous fluid. The time step between two consecutive frames is taken as $\Delta t = 10s$. The initial fiber configuration in numerical simulations is taken from the first experimental frame (see also movies 1 and 2 in supplementary material). $(a)$ Gliding event: parameter values are $Be=200$ and $\xi=7.71$. The gliding event is characterized by a short trapping period around the obstacle followed by a drift motion. $(b)$ Trapping event: parameter values are $Be=210$ and $\xi=9.35$. The trapping event is characterized by a prolonged trapping period of the fiber around the obstacle. $(c-d)$ Evolution of the total velocity of the center of mass of the fiber as function of time in experiments (circles with grey shaded error bars) and numerical simulations (lines) for various depths of the obstacle, respectively for $(c)$ gliding and $(d)$ trapping events. The velocity and the time are scaled respectively by the settling velocity of the corresponding rigid fiber oriented perpendicularly to the direction of gravity, $V_{\perp}$, and the characteristic settling time $T=L\eta/W$.}
    \label{fig: chrono_trapping_gliding_and_velocities}
\end{figure}

The presence of the obstacle leads to two main outcomes that depend on the mechanical and geometrical properties of the system:  the fiber can either  glide  along the obstacle (see Fig.~\ref{fig: chrono_trapping_gliding_and_velocities}$(a)$) or remain trapped around it (see Fig.~\ref{fig: chrono_trapping_gliding_and_velocities}$(b)$). These events results from the complex interplay between internal elastic stresses, contact forces and hydrodynamic interactions with the embedded rigid obstacle. 

Before addressing the role of each of these mechanisms on the fiber trajectory, we characterize these two situations and use them to validate our numerical method.
Fig.~\ref{fig: chrono_trapping_gliding_and_velocities} shows numerical and experimental realizations of the gliding and trapping events for a semi-flexible fiber ($Be\sim 200$)
settling against an obstacle (the complete set of parameters is provided in Appendix \ref{appendix:sets_of_parameters}). 
In the simulations, the initial position of the fiber centerline and its mechanical properties are directly extracted from the experiments. Note that to compute the elasto-gravitational number $Be$ in our multibead approach, we have used the volume of the object as a continuous fiber and not as a chain of beads, which is smaller by a factor 2/3 \cite{marchetti_deformation_2018}. 

As shown by the chronophotographies in panels $(a)-(b)$, the simulations agree qualitatively well with experiments for both events. In addition, the time evolution of the settling speed, reported in panels $(c)-(d)$, shows excellent quantitative agreement, thus confirming the adequacy of our method to capture fiber-obstacle interactions in a viscous fluid. In the gliding case, the fiber starts close to its equilibrium shape with a settling speed larger than a rigid one ($V> V_{\perp} $), (see panels $(a)$ and $(c)$). As it settles, it progressively slows down until it partially hits the obstacle and reaches its minimum speed. The speed increases again as the fiber glides along the obstacle. Upon release, it is more aligned with gravity and reaches its maximum velocity ($V/V_{\perp} \approx 1.2$). Eventually it slowly relaxes to its equilibrium shape: the velocity decreases to its initial value and the fiber drifts sideways as it reorients perpendicular to gravity.   
In the trapping case, the initial lateral distance between the fiber and obstacle, $D_x$, is smaller. As a result the fiber slows down continuously, wraps around the obstacle and finally remains in a trapped configuration indefinitely (see Fig.~\ref{fig: chrono_trapping_gliding_and_velocities}$(b)$ and $(d)$). The decrease in the settling speed is more pronounced when the fiber touches and wraps around the obstacle in the interval $1.3 < t/T < 4$, where $T=L\eta/W$ is the characteristic settling time.
The trapped configuration in the steady regime is asymmetric with respect to the obstacle center of mass. Such asymmetry originates from the tangential contact forces and results from an equilibrium configuration that minimizes the total energy due to external (gravity and contact) and internal elastic forces. 

We note that the settling velocity of the fiber weakly depends on the obstacle depth $D_O$ in numerical simulations, and is close to the experimental value, for which the obstacle is spanning the entire depth of the tank, with typical value $D_O/L\approx 4$ (see Fig.~\ref{fig: chrono_trapping_gliding_and_velocities}$(c-d)$).
This weak dependence can be understood by the fact that the fluid in the tank is at rest and the dominant portion of the flow induced by the fiber decays as $r^{-1}$. Therefore, the response of the obstacle, to maintain its position, is weak for large separation distances. As a result the extremities of the obstacle that are far from the fiber barely affect its motion. That is why we only observe a slight decrease in the velocity of fiber close to the obstacle when $D_O$ increases.  We note that in the case of a fiber transported by a convective, e.g.\ plug or Poiseuille flow, the fiber velocity would be strongly correlated to the obstacle depth as its response to the ambient flow (which is no longer zero) has a magnitude close to the fiber speed.

Now that the gliding and trapping events have been described, and the numerical method validated, we use numerical simulations and analytical tools to explain their mechanical origin, identify their key parameters and to explore their potential for sorting applications.

\subsection{Gliding events: tilting and lateral displacement induced by fiber-obstacle interactions}\label{effects_obstacle}

The gliding motion along the obstacle induces a tilt of the fiber centerline with respect to gravity, measured at the midpoint, denoted $\theta(t)$ (see Fig.~\ref{fig: lateral_shift_numeric_versus_theory}$(a)$). Here, the origin of time ($t=0)$ is taken when the fiber leaves the obstacle. As mentioned earlier, this tilt induces a lateral drift as the fiber reorients back to its horizontal equilibrium shape.   
We define this lateral shift, $ \Tilde{\delta}x=\delta x / L$, as the lateral displacement of the fiber center of mass between a given starting time, here $t=0$, and the final equilibrium state. 

In the elongated limit ($\varepsilon = 2a/L \ll 1$), the velocity and lateral displacement of weakly flexible fibers ($Be\ll1$) with initial orientation $\theta(0)$, and negative curvature, has been computed analytically by Li \textit{et al.} \cite{li_sedimentation_2013} using slender body theory \cite{keller_slender-body_1976, johnson_improved_1980, tornberg_simulating_2004}. Their approach is based on a multiple-scale analysis \cite{hinch_perturbation_1991, bender_advanced_1999} where they identify two relevant independent timescales in the fiber dynamic: the time for the fiber to settle its length $L$, of order  $\mathcal{O}(1)$, and the time to reorient toward its equilibrium configuration, which is much slower and of order $\mathcal{O}(Be^{-1})$. In the case of a fiber with uniform thickness, mass and bending stiffness, they found that the leading order settling velocity is given by $\bm{U}=2c_0\cos(\theta)\bm{\hat{t}}(\theta) - c_0\sin(\theta) \bm{\hat{n}}(\theta)$, where $(\bm{\hat{t}},\bm{\hat{n}})$ are the tangent and normal vector at the midpoint of the centerline, and $c_0 = \ln{(1/\varepsilon^2 e)}$. Therefore, the velocity components in the lab frame, $U_x$ and $U_y$, are obtained by dotting separately with the unit vectors $\bm{\hat{x}}$ and $\bm{\hat{y}}$:

\begin{figure}[ht]
   \centering
    \includegraphics[width=\textwidth]{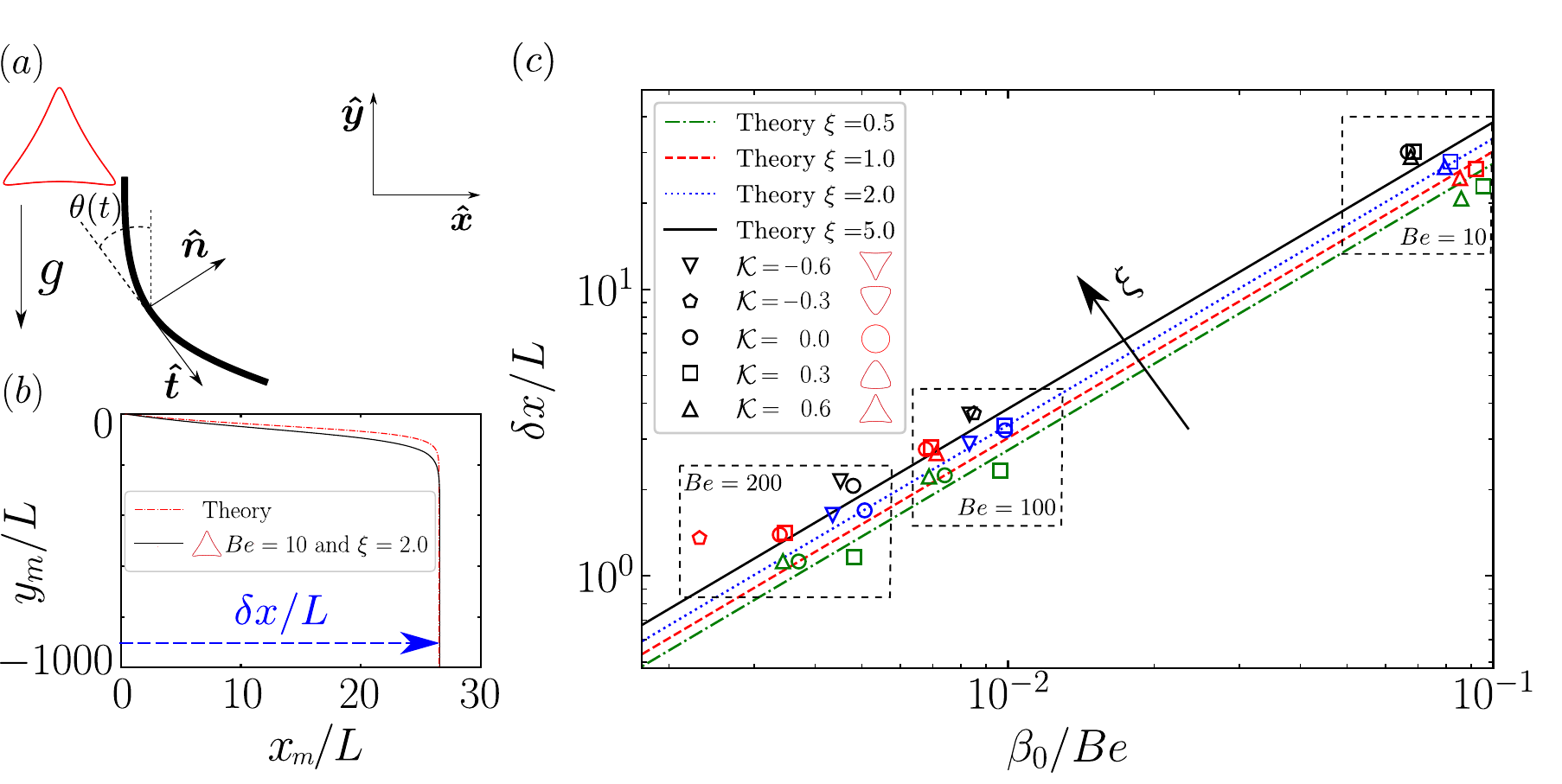}
    \caption{$(a)$ Illustration of a fiber at the edge of the obstacle (downstream). $\hat{\bm{t}}$ and $\hat{\bm{n}}$ are respectively the unit tangent and unit normal vectors. The orientation angle $\theta(t)$ is defined between $\hat{\bm{t}}$ and the direction of gravity $-\hat{\bm{y}}$. $(b)$ Results from numerical simulations (solid line) and from theoretical predictions (dash-dot line) showing the trajectory of the center of mass of the fiber for $Be=10$ and $\xi=2$. The initial configuration of the fiber is considered at the edge of the obstacle (downstream), with an initial orientation $\theta(0)$ induced by its shape, here for $\mathcal{K}=0.6$. $(c)$ Scaled lateral shift $\delta x / L$ versus $\beta_0/Be$. Comparison of the theoretical predictions (lines) with numerical results (symbols) done with one slice on the obstacle, for various values of the geometrical parameter $\xi$ and the conformal mapping parameter $\mathcal{K}$. The corresponding values of $\varepsilon$ are $0.031,0.016,0.008$ and $0.0032$, respectively for $\xi = 0.5, 1, 2 \;\text{and} \;5$.  The showed numerical results (symbols) correspond to three sets of data for $Be = 10, 100$ and $200$. Parameter values: $\Delta s/\mathcal{C}=0.01$, $\alpha=2$ and $D_y/L=5$.}
    \label{fig: lateral_shift_numeric_versus_theory}
\end{figure}

\begin{equation}
   U_x =  \bm{U}\cdot\bm{\hat{x}} = \frac{c_0 h}{1 + h^2}
    \label{eq: velocity_in_x_leading_order},
\end{equation}
\begin{equation}
    U_y = \bm{U}\cdot\bm{\hat{y}} = -c_0\left(1 + \frac{1}{1 + h^2}\right)
    \label{eq: velocity_in_y_leading_order},
\end{equation}
where
\begin{equation}
    h(t) = \tan{(\theta(0))} \exp{((2CBe)t)}
    \label{eq: factor_h_velocity},
\end{equation}
with 
\begin{equation}
    C = \frac{7}{400} + c_0^{-1}\left(\frac{1813 - 300\pi^2 + 630\ln{(2)}}{18000}\right).
    \label{eq: constant_big_c}
\end{equation}

Finally, the lateral shift can be deduced by integrating Eq.\eqref{eq: velocity_in_x_leading_order} and assuming $0 < \theta(0) \leq \pi/ 2$ 
\begin{equation}
    \Tilde{\delta} x = \int_0^{\infty} (\bm{U}\cdot\bm{\hat{x}})\deriv t = \frac{c_0}{2CBe}\beta_0
    \label{eq: lateral_displacement_analytic},
\end{equation}
where 
\begin{equation}
    \beta_0 = \frac{\pi}{2} - \theta(0).
    \label{eq: beta_0}
\end{equation}
Eq.\eqref{eq: lateral_displacement_analytic} is valid in the slender body regime if $C > 0$, i.e.\ when $\varepsilon = 2a/L < 0.196$. Holding $\beta_0$ constant in Eq.\eqref{eq: lateral_displacement_analytic}, we observe that, the lateral shift is inversely proportional to the elasto-gravitational number $Be$ and the larger displacements are achieved for very stiff fibers, $Be \rightarrow 0$. The lateral shift is also proportional to the initial angle $\beta_0$ by holding $Be$ constant. We also notice that, $\Tilde{\delta}x = 0$  when  $\theta(0)=\pi/2$:  a fiber whose unit tangent at the midpoint is initially oriented perpendicular to the direction of gravity, will not experience a drift motion, as expected. 

\begin{figure}
   \centering
    \includegraphics[width=\textwidth]{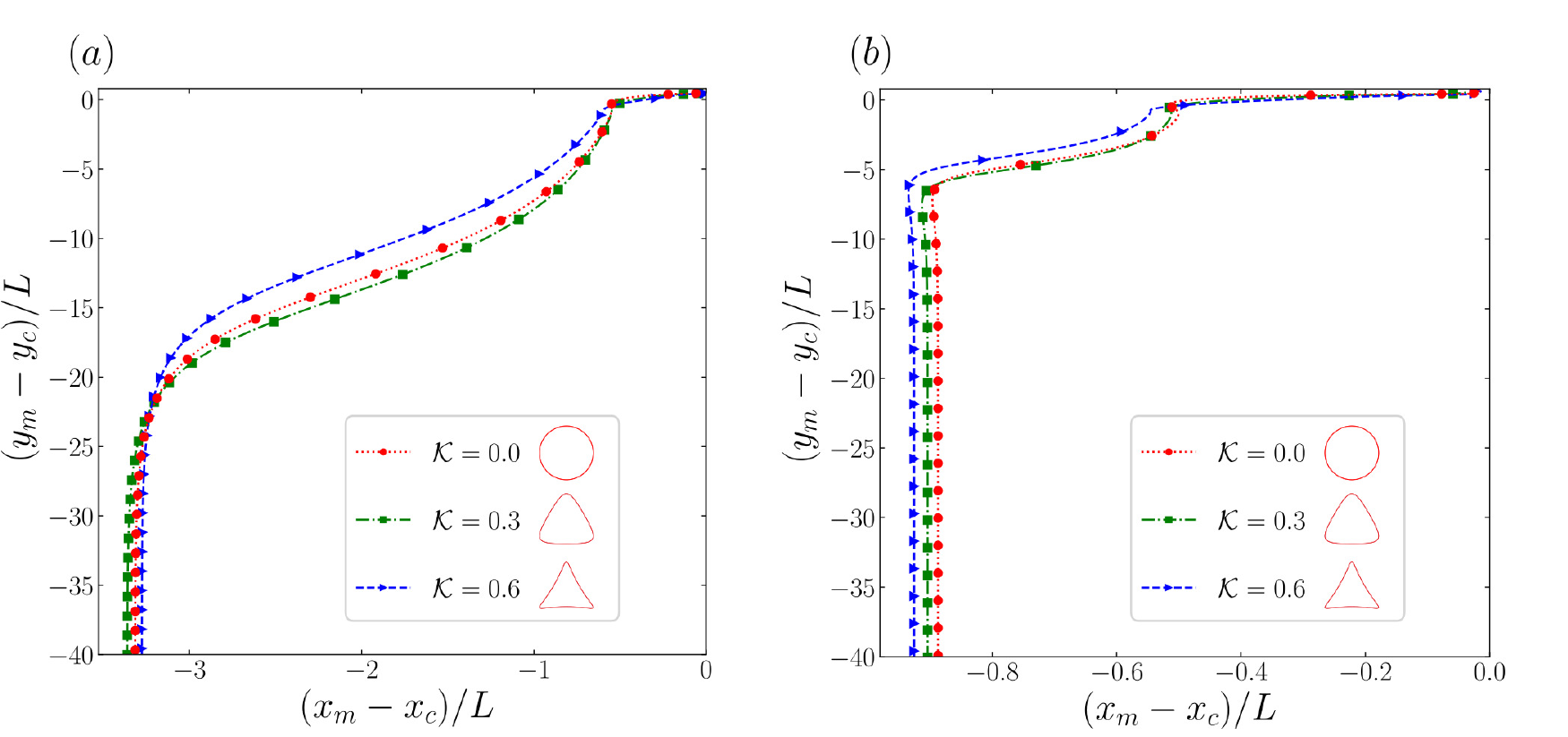}
    \caption{Results from numerical simulations of the trajectory of the center of mass of the fiber for $\xi=1$ and three different shapes of the obstacle ($\mathcal{K}= 0, 0.3$ and $0.6)$. The initial offset of the fiber midpoint is set horizontally to $D_x=0.05L$ with respect to the center of mass of the obstacle. $(a)$ With an intermediate value of the elasto-gravitational number $Be = 100$. $(b)$ With a large value of the elasto-gravitational number $Be=1000$. Note that $x_c$ and $y_c$ represent the position of the center of mass of the obstacle in $xy$ plane. Parameter values are as in Fig.~\ref{fig: lateral_shift_numeric_versus_theory}.}
    \label{fig: lateral_shift_shape}
\end{figure}

Figure ~\ref{fig: lateral_shift_numeric_versus_theory}$(b)$ shows a typical trajectory of the  midpoint of a rigid fiber with $Be=10$, $\xi=2$, after it has hit an upward-pointing triangular obstacle  ($\mathcal{K}=0.6$). The black solid line corresponds to the numerical simulation of the full system and the red dashed line to the theoretical prediction obtained by integrating \eqref{eq: velocity_in_x_leading_order} and \eqref{eq: velocity_in_y_leading_order} in time with the initial tilt angle taken from the numerics (here $\theta(0)=0.38$). While the two curves show some discrepancies at short time when the fiber reorients, probably due to the fact that hydrodynamic interactions with the obstacle are not taken into account in the theoretical approach, their lateral shift matches exactly at long times when equilibrium is reached.  
Fig.~\ref{fig: lateral_shift_numeric_versus_theory}$(c)$ compares the predicted lateral shift in Eq.~\eqref{eq: lateral_displacement_analytic} with numerical simulations for a large range of realizations involving different obstacle shapes $\mathcal{K} = -0.6,\dots,0,6$, relative length $\xi = 0.5,\dots,5$. As before, the initial tilt angle in Eq.~\eqref{eq: beta_0} is taken from the simulations after hitting the obstacle. Since the theoretical approach is valid only for weakly flexible fibers, we used relatively small values of $Be = 10, 100, 200$ in the simulations.  As expected, the simulated lateral shift $\Tilde{\delta}x$ is monotonic in $\beta_0/Be$. Its values is maximal the most rigid case, i.e. the largest $\beta_0/Be$. Similarly to the theoretical prediction we observe a slow increase of $\Tilde{\delta}x$ with $\xi$ (i.e. $\varepsilon^{-1}$) for a given value of $\beta_0/Be$: for the same mechanical properties and a given obstacle shape, elongated fibers exhibit larger displacements than short ones. While theory and numerics agree reasonably well for $\xi > 1$ ($\varepsilon \ll 1$), a slight shift appears for $\xi=5$ (black symbols vs.\ black solid line) due to the fact that the fiber has a positive curvature at its midpoint in the simulations when it leaves the obstacle (see movie 3 in the supplementary material), which violates the hypothesis of negative initial curvature in the theoretical approach. This error is decreased if the initial angle is taken at a later time in the simulations when the curvature at the midpoint becomes negative. For $\xi \leq 1$, the numerical simulations (red and green symbols) present some scatters which reflect the effects of hydrodynamic interactions of order $\mathcal{O}(1/d)$ induced by the obstacle, $d$ being the fiber-obstacle  distance. Note that, these effects are not taken into account in the theoretical approach where the fiber is isolated. In addition to the dependence of the lateral shift on $\xi$, we observe also the dependence of the lateral shift on the initial orientation angle induced by the obstacle shape. Before embarking in a detailed explanation on the dependence of $\Tilde{\delta} x$ on the obstacle shape, we notice that for large values of $\xi$, i.e. $\xi=5$, the lateral shift is quasi-independent of the obstacle shape. For instance, for $\beta_0/Be \approx 10^{-1}$, the induced local deflection along the fiber is lower than in the other cases ($\xi<5$), therefore the initial fiber configurations are quasi-straight and have the same initial orientation angles at the edge of the obstacle (downstream) for all $\mathcal{K}$ (see movies 4-6 in the supplementary material). However, at smaller $\xi$, we observe a slight dependence of $\Tilde{\delta} x$ on the obstacle shape. This dependence is due to the different curvatures of the obstacle sides, i.e. negative curvature ($\mathcal{K}=-0.6$ and $\mathcal{K}=0.6$) or positive curvature ($\mathcal{K}=-0.3$, $\mathcal{K}=0$ and $\mathcal{K}=0.3$), and depends on the fiber flexibility. Finally, the increased discrepancies between simulations and theory at $Be=200$ provide an upper bound for the range of validity of the theory which relies on an expansion at small $Be$.
To include larger values outside the small $Be$ constraint, we perform numerical simulations for $\xi=1$ with $Be=1000$. Typical trajectories for $Be=100$ and $Be=1000$ are presented in Fig. \ref{fig: lateral_shift_shape} for three shapes. In Fig.~\ref{fig: lateral_shift_shape}$(a)$ for $Be=100$, we observe that the lateral shift is maximized where the curvature of the obstacle shape has a positive sign ($\mathcal{K}=0$ and $\mathcal{K}=0.3$) and minimized where it is negative ($\mathcal{K}=0.6$) as observed in Fig.~\ref{fig: lateral_shift_numeric_versus_theory}. However, for a large value of $Be$ ($Be=1000$) shown in Fig.~\ref{fig: lateral_shift_shape}$(b)$, the trend is different, the lateral shift is maximized for $\mathcal{K}=0.6$ compared to the other cases $\mathcal{K}=0$ and $\mathcal{K}=0.3$. This difference can be explained by the fact that for $\mathcal{K}=0.6$, the fiber takes longer to release from the obstacle (characterized by the plateau region) than for $\mathcal{K}=0$ and $\mathcal{K}=0.3$, due its high flexibility which promotes higher adherence to the obstacle curvature.
Nevertheless, for all shapes the angle at which the fiber leaves the obstacle is always small (typically $\theta<\pi/6$) and only slightly depends on the obstacle shape. The obstacle thus always tends to align the fiber with the direction of gravity; from there, the fibers have to reorient towards their equilibrium shape while drifting laterally; the magnitude of the lateral drift then strongly depends on the fiber's length and flexibility through $Be$, which offers opportunities for sorting fibers according to their characteristics.

\begin{figure}
   \centering
    \includegraphics[width=\textwidth]{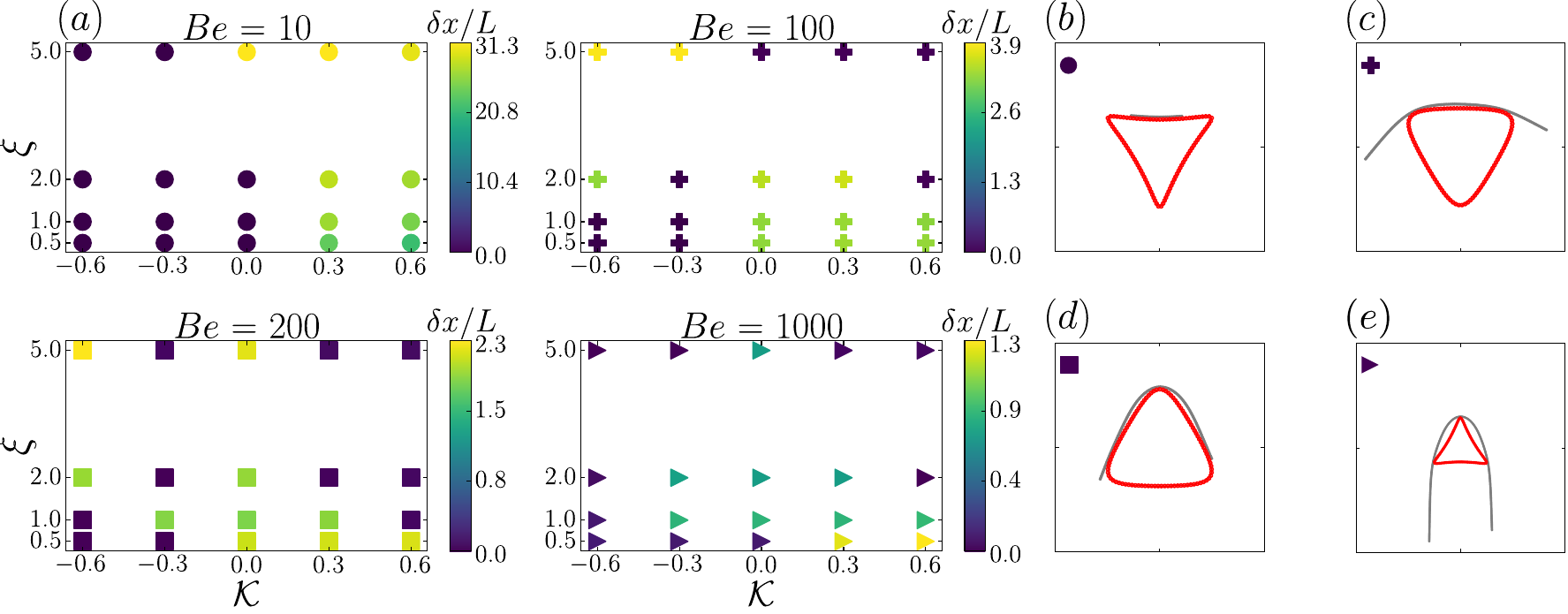}
    \caption{Results from numerical simulations computed with one slice on the obstacle. The initial offset of the fiber midpoint is set horizontally to $D_x=0.05L$ with respect to the center of mass of the obstacle. $(a)$ Phase diagram, showing the scaled lateral shift $\delta x/L$ for various values of the elasto-gravitational number $Be$, the relative length $\xi$ and the conformal mapping parameter $\mathcal{K}$. The purple symbols ($\delta x / L \sim 0$), denote the trapping events. $(b-e)$ Trapping events resulting from the phase diagram (see also movies 7-10 in supplementary material). $(b)$ $Be=10$, $\xi=0.5$ and $\mathcal{K} = -0.6$. $(c)$ $Be=100$, $\xi=2$ and $\mathcal{K} = -0.3$. $(d)$ $Be=200$, $\xi=2$ and $\mathcal{K} = 0.3$. $(e)$ $Be=1000$, $\xi=5$ and $\mathcal{K} = 0.6$. Parameter values are as in Fig.~\ref{fig: lateral_shift_numeric_versus_theory}. }
    \label{fig: lateral_shift_for_different_be}
\end{figure}

Finally, we systematically explore the influence of the parameters on the lateral shift with a large set of simulations in the range $-0.6 \leq \mathcal{K} \leq 0.6$, $10 \leq Be \leq 1000$, $1 \leq \xi \leq 5$, for a given initial lateral offset $D_x/L=0.05$. The results, reported in  Figure ~\ref{fig: lateral_shift_for_different_be}, show the coexistence of the gliding and trapping states, the latter corresponding to the purple symbols ($\delta x / L = 0$). Before undertaking a detailed investigation of the trapping events in Sec. \ref{trapping}, we qualitatively highlight the features of the resulting phase diagram (Fig.~\ref{fig: lateral_shift_for_different_be}$(a)$) for each value of $Be$.
Firstly, we observe that the maximum value of the lateral displacement $\delta x/L$ starts to saturates with  $Be^{-1}$ in the flexible regime $Be\geq 200$,  which clearly indicates where the linear relationship obtained from the theory \eqref{eq: lateral_displacement_analytic} breaks down. 
Secondly, we notice that in the flexible regime ($Be=1000$), for a given obstacle shape $\mathcal{K}$, the lateral shift in the gliding events is inversely proportional to the fiber length $\xi$ (and thus proportional to  $\varepsilon$). This trend contrasts with the rigid and semi-flexible regimes ($10\leq Be \leq 200$) where the opposite is observed and predicted by the theory \eqref{eq: lateral_displacement_analytic}. Indeed as the flexibility increases, local deformations happen near the tip of the longest fibers as they exit the obstacle, which tends to reorient them earlier than shorter ones and therefore reduce their lateral displacement.
Thirdly, trapping states are mostly localized in the bottom left (short fibers on downward-pointing triangles) and upper right (long fibers on upward-pointing triangles) corners of the $(\mathcal{K},\xi$) plane for $Be\geq 100$. In the most rigid regime $Be=10$ trapping only occurs on obstacles with a large incident contact surface ($\mathcal{K}\leq 0$). 

\subsection{Investigation of trapping events}\label{trapping}
As shown in Fig. \ref{fig: lateral_shift_for_different_be}$(b-e)$, for a given initial lateral offset (here $D_x = 0.05L$), the fiber can be trapped in many different ways depending on the obstacle shape $\mathcal{K}$, elasto-gravitational number $Be$, and relative length $\xi$. Some of these configurations might not seem intuitive at first sight and need to be rationalized. The goal of this section is to systematically explore the wide diversity of trapping states in order to connect them to the geometric and mechanical parameters of the system. 

In the absence of surface roughness, steric forces with the obstacle are exclusively directed along the normal of the fiber centerline. In this regime, trapping is only  possible for symmetric configurations i.e.\ with zero lateral offset ($D_x/L=0$), where the gravity forces balance on both sides of the fiber.
However, surface asperities appear both in experiments and simulations: even though the crosslinked fibers are smooth, the 3D-printer has a finite resolution and defects might occur, while the simulated obstacles and fibers have a maximum interpenetration length $\delta/a\approx 0.17$ (see Section \ref{sec:numerical_method}).
Surface roughness generates friction directed along the fiber centerline. These tangential forces can balance asymmetric gravity forces and thus prevent the fiber from slipping away from the obstacle.  In numerical simulations, friction forces correspond to the tangential component of the steric forces between obstacle and fiber beads $F^R_{\tau}$.
The trapping probability and trapping configurations are obviously highly sensitive to the details of the surface roughness. However, we did not try to match the experimental roughness in the simulations. The focus of this section is, for a given roughness value, to understand the mechanisms that lead to the observed trapping states.

\begin{figure}
   \centering
    \includegraphics[width=0.75\textwidth]{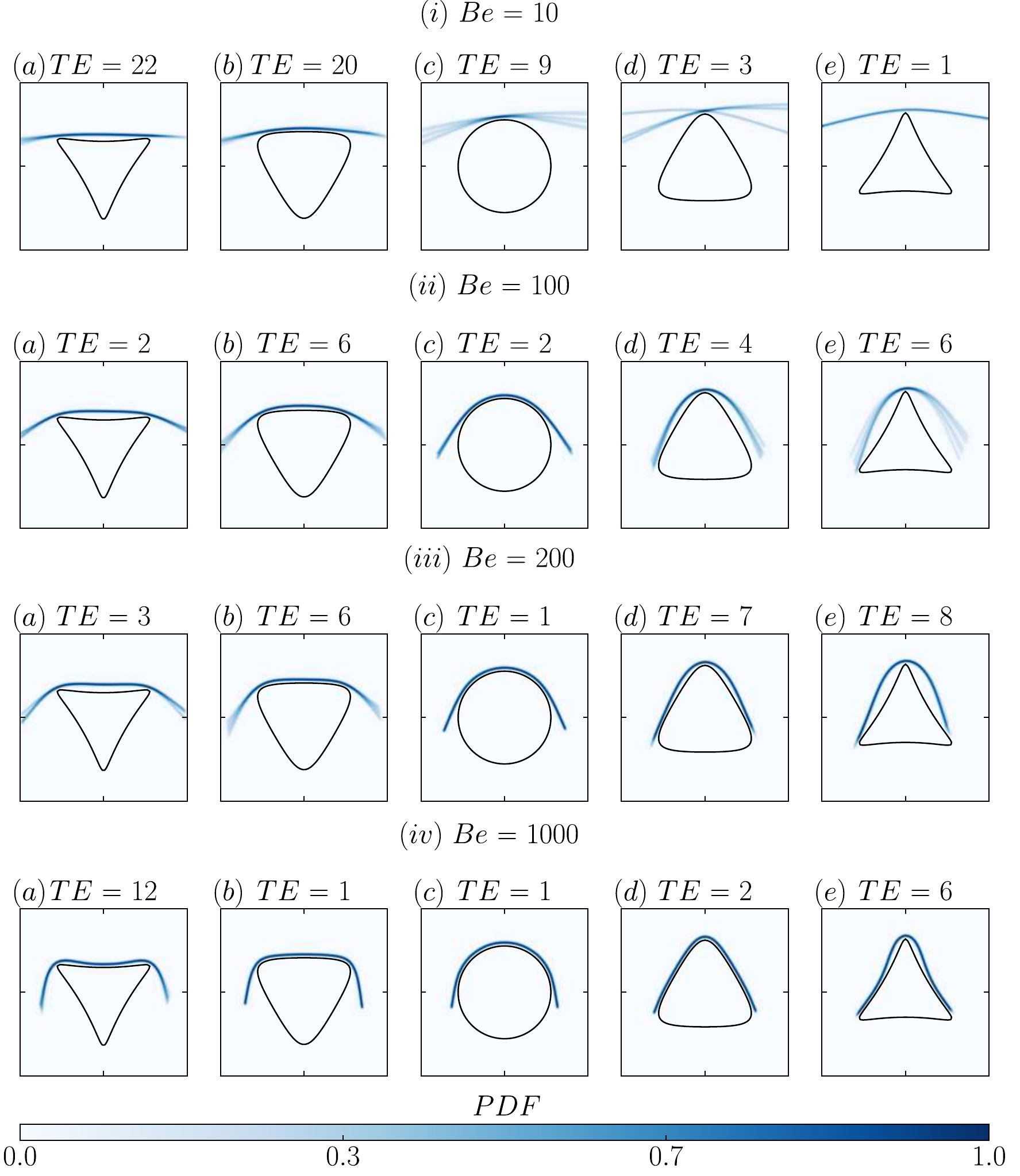}
    \caption{Probability distribution of the trapping configurations of the fiber around the obstacle, for $\xi=2$ and various values of $Be$ and $\mathcal{K}$. $TE$ stands for the number of trapping events. Note that we used $10^2$ initial configurations per shape and for a given value of $Be$, generated by varying the initial offset between $-0.5$ and $0.5$ with a step size of $10^{-2}$. For the sake of clarity, the initial offsets of the showed configurations fall between 0 and 0.5, since the remaining configurations are the same by symmetry. Parameter values are as in Fig.~\ref{fig: lateral_shift_numeric_versus_theory}. 
    }
    \label{fig: pdf_trapping_events}
\end{figure}

To do so, we systematically explored the four-dimensional parameter space ($\xi, \mathcal{K}, Be, D_x/L$). For each simulation, the fiber is released at a fixed height $D_y/L = 5$  with its equilibrium shape and the initial lateral offset is varied between $D_x/L = \pm 0.5$  with steps of size $0.01$. Figure \ref{fig: pdf_trapping_events} shows the probability density function (PDF) of finding the trapped fiber centerline, with relative length $\xi=2$, as a function  of the elasto-gravitational number $Be$ and obstacle shape $\mathcal{K}$. 
Owing to the symmetry of the problem, the PDF is computed only for the trapping events (TE in the figure) happening in the range $D_x/L = [0;0.5]$, which represents 51 simulations for each panel.

\begin{figure}
   \centering
    \includegraphics[width=\textwidth]{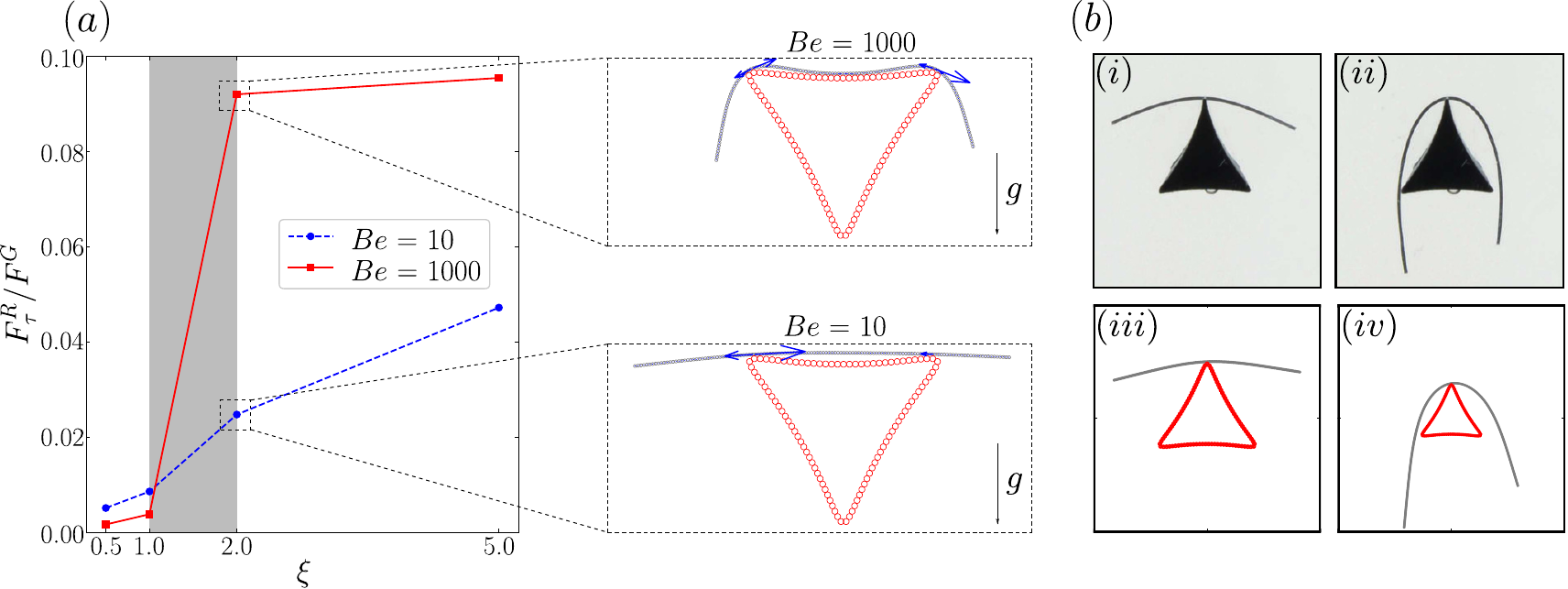}
    \caption{$(a)$ Evolution of the tangential component (frictional force) of the steric force scaled by the gravitational force for two different values of $Be$ and various values of $\xi$, $D_x=0.05L$ and $\mathcal{K}=-0.6$. The close-up views show the corresponding distributions of the tangential component of the steric force along the fiber centerline, for $\xi=2$. $(b)$ Trapping events from experimental $(i-ii)$ and numerical $(iii-iv)$ results obtained for $\mathcal{K}=0.6$. $(i)$ $Be=36$ and  $\xi=2$. $(ii)$ $Be=210$ and $\xi=4$. $(iii)$ $Be=10$ and $\xi=2$. $(iv)$ $Be=200$ and $\xi=5$.}   
    \label{fig: tangential_component_and_trapping_config}
\end{figure}

In the stiff limit, $Be=10$ (Fig.\ \ref{fig: pdf_trapping_events}$i$), the fiber barely deforms and cannot fit the obstacle shape.  The trapping configurations result from the balance between the tilt induced by the lever arm due to gravity and the friction forces at the contact points with the obstacle. The trapping probability of the fiber therefore decreases as the tip of the obstacle narrows, as shown by the  drop in the number of trapping events (TE) between $\mathcal{K}=-0.6$ and $\mathcal{K}=0.6$ (panels $i(a-e)$).
This competition between lever arm and friction at the tips obstacle leads to  surprising trapping states for upward pointing triangles $i(d-e)$. 
In the highly flexible regime, $Be=1000$  (Fig.\ \ref{fig: pdf_trapping_events}$iv$), the fiber better fits the obstacle shape and its freely hanging extremities are aligned with the gravitational field.  Such alignment with gravity induces a tangential load along the fiber centerline, which competes with  friction in the high curvature regions of the obstacle. As a result, the fiber is mostly trapped by obstacles with high curvatures, i.e.\ for $\mathcal{K} = \pm 0.6$ (panels $iv(a)$ and $iv(e)$), and easily slips away from smooth obstacles (panels $iv(b-d)$).\\
To further quantify the main differences between the stiff and flexible limits, we measure the total tangential (i.e.\ frictional) component of the resultant steric force, $F^R_{\tau}$, acting on the fiber against an obstacle with a high trapping probability, $\mathcal{K}=-0.6$, as a function of the relative length $\xi$, for a small initial offset $D_x/L=0.05$ (see Fig. \ref{fig: tangential_component_and_trapping_config}$(a)$).
First we notice that, regardless of the rigidity, the friction force is small for short relative lengths $\xi\leq 1$, where the fiber fully rests on the  base of the triangle.  When $\xi>1$, the rigidity makes a big difference since the contact line with the obstacle becomes more and more localized along the fiber. In the rigid case ($Be=10$), the tilt induced by the lever arm effect increases with $\xi$ and so does the friction force to counterbalance the resulting tangential motion. In the flexible regime ($Be=1000$), as soon as the fiber extremities stick out of the obstacle ($\xi>1$), they align with gravity and generate a strong tangential load  (i.e.\ a pulley effect). As a result the friction force  jumps suddenly, by a factor 24, between $\xi=1$ and $\xi=2$ and is at least twice larger than the friction on the rigid fiber. This is illustrated for $\xi=2$ in the close-up views of Fig. \ref{fig: tangential_component_and_trapping_config}$(a)$, where the larger compliance of the flexible fiber leads to an extra frictional contact near the right edge of the triangle compared to the rigid one.  For  longer flexible fibers ($\xi=5$), the curvature  near the edges of the obstacle saturates and so does the friction force. 

In the intermediate semi-flexible regime ($Be = 100, \, 200$, see Fig.\  \ref{fig: pdf_trapping_events}$ii-iii$), the increasing flexibility promotes alignment of the extremities with gravity but the fiber cannot fit obstacles with high curvatures. As a result fibers are much less trapped for $\mathcal{K}=-0.6$ because the strong tangential motion induced by the hanging extremities beats the friction with the pointy vertices of the triangle, while the increased contact area with a smooth inverted triangle, $\mathcal{K}=-0.3$, allows for more trapping configurations. For the upward pointing triangles, the fiber can bend enough so that friction at the tip can beat tangential motion. Fiber flexibility also allows one of its extremities to touch one of the sides of the triangle. The vertical component of the steric force between the obstacle side and the fiber extremity, which acts against gravity, increases as the obstacle curvature becomes negative, leading to higher trapping probability for $\mathcal{K}=0.6$ than for $\mathcal{K}=0.3$. 

Some of the exotic, asymmetric, trapping states observed in numerical simulations with non-circular obstacles have also been reported in experiments.  Figure \ref{fig: tangential_component_and_trapping_config}$(b)i-ii$ shows two occurrences of relatively long fibers being trapped on a curved triangle  ($\mathcal{K}=0.6$) in the stiff ($Be\approx36$ and $\xi=2$) and semi-flexible ($Be\approx210$ and $\xi=4$) regimes. The two panels below (Fig.\ \ref{fig: tangential_component_and_trapping_config}$(b)iii-iv$) show qualitatively similar trapping states from numerical simulations in the explored parameter state ($Be=10$ and $\xi=2$ for the stiff regime and $Be=200$ and $\xi=5$ for the semi-flexible case), thus confirming  the relevance of our parametric exploration, and more importantly, showing that even a small amount of friction can lead to nontrivial trapping configurations. This good agreement with experiments suggests that the effective friction coefficient in our simulations is at least equal or larger than the experimental one.
Finally, for the range of parameters considered, our analysis shows that trapping is overall minimized with smooth obstacles ($\mathcal{K}=0,0.3)$. These findings could be used in the design of fiber sorting devices where trapping or long residence times are undesired.

\subsection{Toward a sorting device}

We have shown that an obstacle can reorient a settling fiber, and thus deviate its trajectory. This reorientation and its subsequent lateral motion strongly depend on the geometrical and mechanical properties of the fiber ($Be$, $\xi$) and  the obstacle shape ($\mathcal{K}$). 

\begin{figure}
   \centering
    \includegraphics[width=0.84\textwidth]{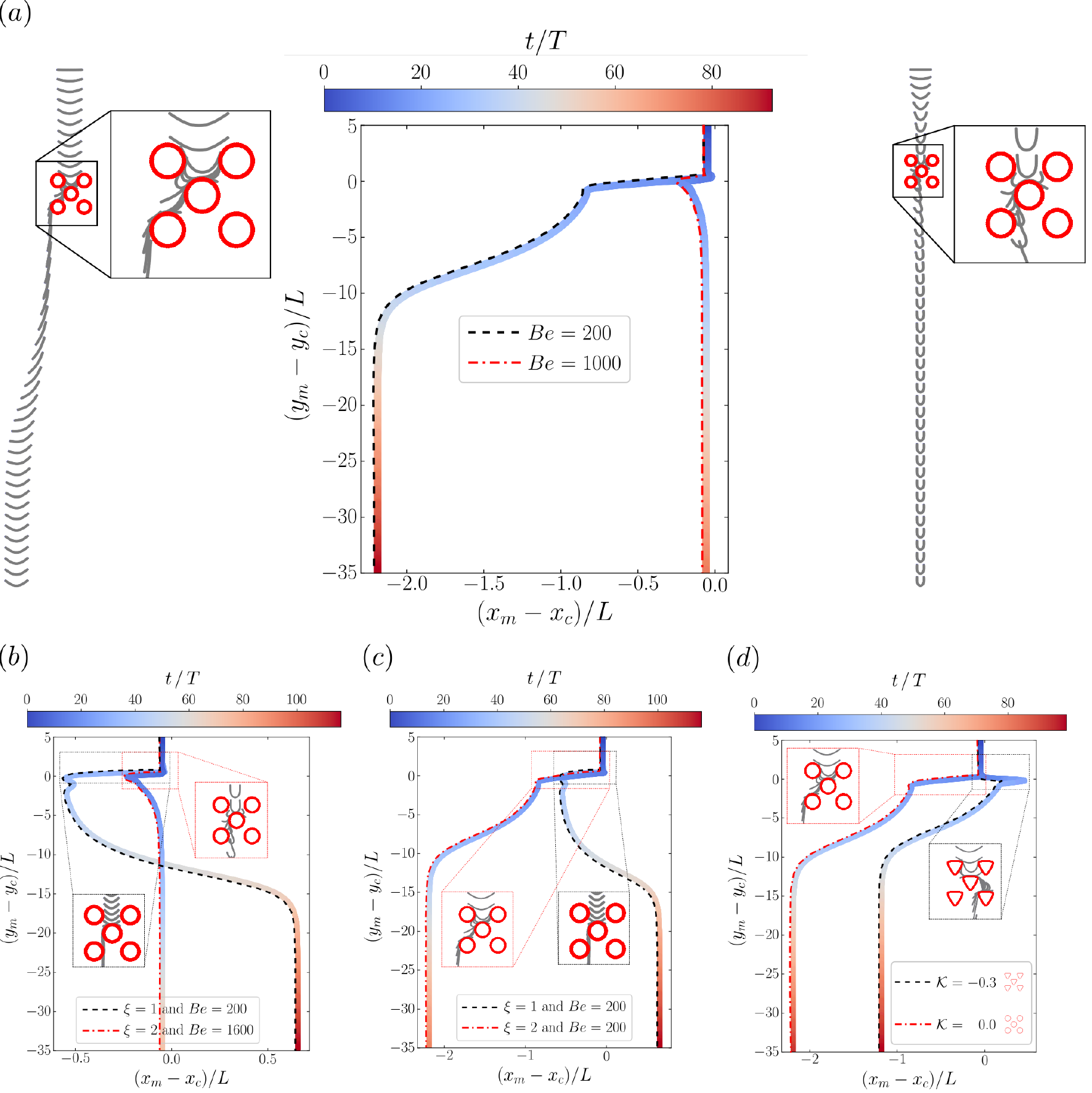}
    \caption{$(a)$ The central panel shows the trajectories of the center of mass of two fibers with the same length $\xi=2$ and different elasto-gravitational number, respectively $Be=200$ and $Be=1000$. The side to side panels show the corresponding chronophotographies of the fiber centerline settling through the unit cell, respectively on the left and right of the central panel (see also movies 11 and 12 in supplementary material). $(b)$ Trajectories of the center of mass of two fibers with the same rigidity $EI$ and different length, respectively $\xi=1$ and $\xi=2$. $(c)$ Trajectories of the center of mass of two fibers with the same elasto-gravitational number $Be=200$ and different length, respectively $\xi=1$ and $\xi=2$. $(d)$ Trajectories of the center of mass of two fibers with the same elasto-gravitational number $Be=200$ and length $\xi=2$, settling through two different unit cells, respectively $\mathcal{K}=-0.3$ and $\mathcal{K}=0$. Parameter values: $\Delta s/\mathcal{C}=0.02$, $\alpha=2$ and $D_y/L=5$.}
    \label{fig: sorting_device}
\end{figure}

We now discuss how to leverage and guide such lateral dispersion with obstacles arranged in a lattice  to sort fibers according to their flexibility and/or length. Our goal here is to provide the proof of concept of a new, passive, sorting strategy rather than a systematic parametric exploration or optimization of the system.
In the following, we analyze the motion of a fiber settling towards a unit cell of a centered square lattice of circular obstacles ($\mathcal{K}=0$). The lattice spacing between nearest obstacles is $d=1.5\omega$, which leaves a gap of $d-\omega=0.5\omega$ for the fiber to go through. The fiber is initially released horizontally with a lateral offset $D_x=0.05L$ with respect to the center obstacle.
We consider two different sorting strategies: 1) sorting fibers by rigidity $EI$ for a given length $L$, 2) sorting by length for a given rigidity.

To showcase the first sorting strategy, we consider two fibers with the same relative length $\xi=2$ that differ by a factor 5 in their rigidity, so that their elasto-gravitational number are $Be=200$ and $Be=1000$ respectively.
Figure\ \ref{fig: sorting_device}$(a)$ shows their conformation and trajectory over time. The most flexible fiber ($Be=1000$) approaches the lattice with an equilibrium  ``U" shape such that its end-to-end width is smaller than the spacing between the obstacles in the first row of the lattice. As a result it directly falls onto the center obstacle,  zigzags around it and escapes from the center of the lattice.
Due to its high elasto-gravitational number, the fiber quickly readjusts to find its equilibrium ``U" shape with almost no lateral displacement ($\delta x/L = 0.01$).
On the other hand, the more rigid fiber approaches the lattice with a ``V" shape that is wider than the entrance. It therefore interacts with the first two obstacles and migrates to the left side of the lattice. It thus lands on the outer side of the left obstacle, allowing it to exit vertically from the diagonal of the lattice 
and slowly reorient back to its equilibrium shape while drifting leftward. As a result, the fibers are separated by a distance $2.13L$ after passing only through one unit cell of the lattice, which clearly illustrates the potential efficiency of the first strategy with a larger lattice.

In the second scenario, only the fiber length changes. The shortest one has a relative length $\xi=1$ and an elasto-gravitational number  $Be=200$. The second one is twice longer ($\xi=2$), and thus has an elasto-gravitational number 8 times higher ($Be=1600$). Due to its small size, the short fiber directly lands on the center obstacle (see Fig.\ \ref{fig: sorting_device}$(b-c)$). After wrapping around it, it aligns with gravity, and exits from the center of the lattice. It finally slowly reorients while drifting sideways which results in a lateral displacement of $\delta x/L = 0.7$. The second fiber is twice longer and 8 times more deformable, which allows it to bend between the pores (see Fig.\ \ref{fig: sorting_device}$(b)$). It therefore follows a very similar trajectory to the first scenario ($Be=1000$ and $\xi=2$) shown in Fig.\ \ref{fig: sorting_device}$(a)$, where very little lateral displacement ($\delta x/L = 5\cdot 10^{-4}$) is observed due to the short reorientation time. 
 If the fibers were rigid regardless of their length, (i.e.\ $EI\gg 1$ so that $Be<1$ for all $L$), then they would barely deform and the separation would be more efficient because the pore size would differentiate them: short fibers ($L/d<0.5$) would fall vertically through the pores while long fibers ($L/d>0.5$) would travel along the diagonals.
However, our simulations suggest that  non-rigid fibers tend to fall vertically and exit from the same location in the lattice regardless of their length. Indeed,  long fibers  ($\xi>1$) are much more deformable than short ones (because of the $L^3$ factor in the elasto-gravitational number $Be$), and  therefore bend and squeeze between the pores to fall in a zig-zag motion. 
Even though these fibers are not sorted inside the network, the lattice is still useful to reorient them so that the difference in their reorientation dynamics, and the subsequent lateral drift, separates them in the obstacle-free area underneath. 

Although in most practical cases, the fibers would be sorted by length and/or rigidity, i.e. with coupled variations in both $\xi$ and $Be$, we can isolate the mechanisms at play by varying $\xi$ and keeping $Be$ constant (i.e. by adjusting the rigidity $EI$) in order to identify the effect of changing the fiber length only. In that case, sorting becomes easier. Indeed, Fig.\ \ref{fig: sorting_device}$(c)$ shows the trajectory of two fibers of different length ($\xi=1$ and $\xi=2$ resp.) with the same elasto-gravitional number $Be=200$. Because the longest fiber has the same deformability as the small one, it is now rigid enough to travel along the diagonal of the lattice. The final separation, here $2.84L$, is four times  greater than in the previous scenario. 


Finally, one may wish to optimize sorting by tuning the obstacle shape. In Figure \ref{fig: sorting_device}$(c)$ we compare the trajectory of a semi-flexible fiber ($Be=200$ and $\xi=2$) crossing  a unit cell made of circles (already analysed above) and of smooth inverted triangles ($\mathcal{K} = -0.3$). The  triangular obstacles have two main effect on the fiber trajectory: 1) the fiber follows a much more complex path across the lattice, leading to a different exit location (through the center instead of the diagonal); 2) due to prolonged contacts with the flat base of the triangles, the residence time of the fiber is increased by $50\%$ ($15T$ instead of $10T$ for  $\mathcal{K}=0$), as expected from our analysis in Section \ref{trapping}. These effects may not be particularly desirable or easy to comprehend for the design of an efficient sorting platform. Furthermore, obstacles with circular cross-sections also minimize fiber trapping, and may thus provide a simple, easy to manufacture solution for optimal sorting.

\section{Conclusions}

The dynamics of a flexible fiber sedimenting in a structured medium is dictated by complex and intricate couplings, combining long-ranged hydrodynamic interactions,  internal elastic stresses and contact forces (steric and/or friction) between the fiber and the surface of obstacles of arbitrary shape. With a combination of theory, numerical simulations and experiments, all of which show excellent quantitative agreement, we explain how these various mechanisms lead to lateral migration or trapping by obstacles. 

Indeed, in a large tank, in the absence of obstacles, flexible fibers adopt an horizontal, curved, equilibrium shape and settle vertically. Upon hitting an obstacle, a fiber changes its orientation by an amount that depends on the obstacle shape $\mathcal{K}$, relative length $\xi$, fiber deformability $Be$ and initial lateral offset $D_x$. As it relaxes back to its equilibrium shape the fiber drifts laterally due to its elongated shape. In the limit of small $Be$, the relaxation time scale and the resulting lateral displacement are inversely proportional to $Be$. This obstacle-induced lateral displacement under gravity can be used to passively sort fibers according to their $Be$ and thus their mechanical properties $E$ and/or length $L$. Preliminary simulations show that fibers can indeed be efficiently sorted by length and/or rigidity in a lattice of circular obstacles. This system has the major advantage of being fully passive, since no energy is needed, low-tech and simple to implement.  The design of a sorting device only requires a tank and 3D-printed obstacles. However, even a small amount of friction (due to roughness) between the fiber and obstacle surfaces can lead to the trapping of the fiber. The trapping configurations and their likelihood depend on the obstacle shape and $Be$, and we find that the  obstacle shape that best avoids trapping is a circular cross-section. 

When transported by a background flow, a flexible fiber can migrate across the streamlines  by dynamically changing its shape in the velocity gradients induced by the obstacle. However, even if the fiber undergoes drastic deformations (buckling, rotation, coiling, snaking,...), the resulting lateral displacement remains small compared to the fiber length ($\delta x/L<1$) \cite{vakil_flexible_2011}. In our system, the reorientation induced by the obstacle generates a lateral shift one to two orders of magnitude larger ($\delta x/L \sim O(1-100)$). We therefore believe our approach could be a complementary and promising alternative to the traditional deterministic lateral displacement (DLD) methods used to sort elongated  particle in microfluidics \cite{mcgrath2014deterministic,salafi2019review}.
In some practical situations (wastewater treatment, textile and micro-plastic clean-up, separation of pathogen populations), and in order to increase the throughput of our method, a whole fiber suspension would be injected in the device. If the fiber concentration is high enough, fiber-fiber hydrodynamic and contact interactions would probably affect their lateral displacement and their trapping likelihood. In the absence of obstacles, sedimenting fiber suspensions are known to exhibit clustering and large scale density fluctuations \cite{saintillan_smooth_2005,gustavsson_gravity_2009,guazzelli_fluctuations_2011,manikantan_effect_2016,schoeller_methods_2021, du_roure_dynamics_2019}. The effect of obstacles and porosity on these collective effects is an open fundamental question that we will tackle in the future to further develop our sorting strategy.

\begin{acknowledgments}
B.D.\ and U.M.\ acknowledge support from the French National Research Agency (ANR), under award ANR- 20-CE30-0006. B.D.\ also thanks the NVIDIA Academic Partnership program for providing GPU hardware for performing some of the simulations reported here.
\end{acknowledgments}

\appendix

\section{Area-preserving conformal mapping}
\label{appendix:conformal_mapping}
We use Riemann area-preserving map of the unit disk \cite{avron_optimal_2004, alonso-matilla_transport_2019} to generate a class of cross-section shapes of the obstacle

\begin{equation}
    \mathcal{F}(z) = \mathcal{W}z + \frac{\mathcal{Y}}{z} + \frac{\mathcal{K}}{\sqrt{2}z^2},
    \label{eq: conformal_map}
\end{equation}
where $z = e^{i\theta}$ with $\theta \in [0, 2\pi]$, $\mathcal{Y}$ and $\mathcal{K}$ respectively  control the aspect ratio and the fore-aft asymmetry of the shape.

The shape parameter $\mathcal{W}>0$ is defined as
\begin{equation}
    \mathcal{W} = \sqrt{1 + \mathcal{Y}^2 + \mathcal{K}^2}.
    \label{eq: shape_parameter_w}
\end{equation} 

Therefore, the coordinates of a given point in the physical $xy$ plane derive from the mapping function \eqref{eq: conformal_map} 
\begin{equation}
   \begin{dcases}
     x = (\mathcal{W} + \mathcal{Y})\cos{\theta} + \frac{\mathcal{K}}{\sqrt{2}}\cos{2\theta}\\
     y = (\mathcal{W} - \mathcal{Y})\sin{\theta} - \frac{\mathcal{K}}{\sqrt{2}}\sin{2\theta}
\end{dcases}
\end{equation}
In our study, we take $\mathcal{Y}=0$ and the resulting geometries are homogeneously dilated in order to have a constant obstacle width $\omega=1$ for all the simulations.

\section{Details on the numerical method}
The surface of the obstacle is discretized in slices  with $N_O$ beads of radius $a_O = \alpha a$, where $\alpha>1$. 
The contour of a slice is discretized with  beads  separated by a distance $\Delta s = 2a_O$ along the perimeter $\mathcal{C}$. 
The slices are then stacked on top of each other to reach the desired obstacle depth. For instance, circular obstacles are built with a hexagonal close-packing arrangement of rings so that the depth of the obstacle is given by $D_O = \sqrt{3}a_O(N_{rings}-1) + 2a_O$, where $N_{rings}$ is the number of rings (i.e.\ slices).

The total number of beads in the system is $N=N_F + N_O$  and $\bm{R} = [\bm{r}_1,\cdots, \bm{r}_{N_F}, \bm{r}_{N_F+1}, \cdots,  \bm{r}_{N}]$ and we denote $\bm{U} = [\bm{U}_1,\cdots, \bm{U}_{N_F}, \bm{U}_{N_F+1}, \cdots,  \bm{U}_{N}]$  the $3N$ vectors collecting their positions and translational velocities.

Since the obstacle is not moving, the obstacle beads must have zero velocity, which can be written as
\begin{equation}
 \bm{\mathcal{J}} \cdot \bm{U} = \bm{0}
 \label{eq: kin_condition}
\end{equation}
where $\bm{\mathcal{J}} = \begin{bmatrix} \bm{0} & \bm{I}\end{bmatrix}$ is a block matrix of size $3N_O \times 3N$, with $\bm{0}$ being a $3N_O \times 3N_F$ zero matrix and $\bm{I}$ the $3N_O \times 3N_O$ identity matrix. The role of $\bm{\mathcal{J}}$ is to select the obstacle beads to which the kinematic conditions apply.

Using the principle of virtual work, one can define the constraint forces associated to these kinematic constraints
\begin{equation}
\bm{F}^{C} = \bm{\mathcal{J}}^{T} \cdot \bm{\lambda}
\end{equation}
where  $\bm{\lambda}$ is a $3N_O$ vector of Lagrange multipliers, homogeneous to a force, exerted on the obstacle beads in order to satisfy the zero velocity constraint.

Because the Reynolds number associated with the fiber motion is relatively small ($Re \ll 1$), the equation of motion of the beads, in the absence of background flow, is given by the linear mobility relation between forces and velocities
\begin{equation}
 \frac{d\bm{R}}{dt} \equiv \bm{U} = \bm{\mathcal{M}} \cdot \left(\bm{F}^{C} + \bm{F}^{\prime} \right)= \bm{\mathcal{M}} \cdot \left(\bm{\mathcal{J}}^{T} \cdot \bm{\lambda} + \bm{F}^{\prime} \right)
\label{eq: mob_problem},
\end{equation}
where  $\bm{F}^{\prime} = \bm{F}^G/N_F + \bm{F}^E + \bm{F}^R$ sums the gravitational and elastic forces acting on the fiber beads as well as the  repulsive contact forces between obstacle and fiber beads. Note that the gravity force acting on a fiber bead is the total fiber weight divided by its number of beads $N_F$. Finally $\bm{\mathcal{M}}$ is the bead mobility matrix  which encodes their hydrodynamic interactions (HI). For convenience we break it into four-blocks: 
    \begin{equation}
    \bm{\mathcal{M}} =
    \begin{bmatrix}
    \bm{M}_{FF} & \bm{M}_{OF}  \\
    \bm{M}_{FO} & \bm{M}_{OO} \\
    \end{bmatrix}
    \label{eq: mob_matrix}.
    \end{equation}
where $F$ and $O$ represent the fiber and obstacle beads respectively. 
Together Eqs. \eqref{eq: kin_condition}-\eqref{eq: mob_problem} form a constrained mobility problem for the Lagrange multipliers $\bm{\lambda}$ and bead velocities $\bm{U}$. After some simple linear algebra, we obtain
 \begin{equation}
    \bm{\mathcal{J}} \cdot \bm{\mathcal{M}} \cdot \bm{\mathcal{J}}^{T} \cdot \bm{\lambda} = -  \bm{\mathcal{J}} \cdot \bm{\mathcal{M}} \cdot \bm{F}^{'}
    \label{eq: con_relation},    
    \end{equation}
which, given the structure of $\mathcal{J}$, simplifies to 
 \begin{equation}
 \bm{\lambda} = - \bm{M}_{OO}^{-1} \cdot  \bm{\mathcal{J}} \cdot \bm{\mathcal{M}} \cdot \bm{F}^{'}
 \label{eq: lag_multipliers}.
\end{equation}
where $\bm{M}_{OO}^{-1}$ is the $3N_0\times 3N_0$ inverse of the mobility matrix of the obstacle beads.
Since the obstacle is not moving over time, $\bm{M}_{OO}^{-1}$ never changes so that it can be precomputed before the time loop.

To sum up, our numerical method is as follows:
\begin{itemize}
    \item Initialize the bead positions and precompute $\bm{M}_{OO}^{-1}$ exactly.
    \item For each time step in the time loop
    \begin{enumerate}
        \item Compute the internal and external forces $\bm{F}^{\prime}$ (see next subsections),
        \item Compute the Lagrange multipliers $\bm{\lambda}$ using \eqref{eq: lag_multipliers},
        \item Compute the bead velocities $\bm{U}$ using  \eqref{eq: mob_problem},
        \item Integrate the bead positions in \eqref{eq: mob_problem} with an implicit time integration scheme to handle stiff filaments, here based on backward differentiation formula (BDF) with adaptative time-stepping \cite{brown_vode_1989}.
    \end{enumerate}
\end{itemize}
In our simulations, the matrix vector products $\bm{M}_{OO}^{-1} \cdot (\cdot)$ and $\bm{\mathcal{M}} \cdot (\cdot)$ in \eqref{eq: lag_multipliers} and \eqref{eq: mob_problem} are computed with Graphic Processing Units (GPU's), which allows to handle large number of beads with a low computational cost and to achieve  linear scaling up to $N\approx 10^4$ \cite{usabiaga_hydrodynamics_2016}.

\subsection{Internal elastic forces}

Mechanical interactions between the fiber beads are governed by forces derived from the total elastic potential, which includes the stretching and bending free energies.  Since we are considering plane deformations, twisting of the fiber is neglected.
The discretized stretching free energy of a fiber made of  $N_F$ beads with $N_F - 1$ links is given by \cite{marchetti_deformation_2018}
\begin{equation}
    E^{S} = \frac{S}{4a}\sum_{i=2}^{N_F}\left(|\bm{t}_i| - 2a\right)^2
    \label{eq: str_energy},
\end{equation}
\par
\noindent
where the tangential vector is defined as $\bm{t}_i=\bm{r}_i - \bm{r}_{i-1}$, $S =\pi E a^2$ is the stretching modulus with $E$ being the Young's modulus. 
On the other hand, the bending free energy in its discretized form is given by  \cite{marchetti_deformation_2018}
\begin{equation}
    E^{B} = \frac{
    B}{2a}\sum_{i=2}^{N_F - 1} \left(1 - \hat{\bm{t}}_{i+1}\cdot\hat{\bm{t}}_{i}\right)
    \label{eq: ben_dis_energy}.
\end{equation}
\par
\noindent
where $B= EI =\pi E a^4/4$ is the bending modulus,  and $\hat{\bm{t}}_i$ the unit tangential vector, $\hat{\bm{t}}_i = \bm{t}_i/|\bm{t}_i|$.
The total elastic potential $H = E^S  + E^B$ is  the sum of the stretching and bending free energies. The internal forces are obtained by taking the gradient of $H$ with respect to the bead positions: $\bm{F}^E = -\nabla_{\bm{R}}H$.

\subsection{Hydrodynamic interactions}

The $3 \times 3$ block $\bm{\mathcal{M}}_{ij}$ of the translational-translational mobility matrix $\bm{\mathcal{M}}$ is given by the Rotne-Prager-Yamakawa (RPY) tensor \cite{wajnryb_generalization_2013, zuk_rotneprageryamakawa_2014}, which approximates the hydrodynamic interactions between two beads $i$ and $j$ centered at $\bm{r}_i$ and $\bm{r}_j$ with radii $a_i$ and $a_j$ 
\begin{equation}
     \bm{\mathcal{M}}_{ij} = \left.\left( \bm{I} +  \frac{a_i  }{6} \nabla^{2} \right)\left( \bm{I} +  \frac{a_j  }{6} \nabla^{2} \right) \mathbb{G}(\bm{r}=\bm{r}' - \bm{r}'')\right|_{\bm{r}'=\bm{r}_i}^{\bm{r}''=\bm{r}_j}
    \label{eq: rpy},
\end{equation}
\par
\noindent
where $\mathbb{G}$ is the Green's function for the steady Stokes problem in an unbounded domain
\begin{equation}
    \mathbb{G}(\bm{r}) = \frac{1}{8 \pi \eta r} \left( \bm{I} + \frac{\bm{r}\otimes \bm{r}}{r^2}\right)
    \label{eq: oseen_tensor},
\end{equation}
with $\eta$  the fluid viscosity and $r = |\bm{r}|$.

The diagonal block is defined as the Stokes drag on a translating sphere 
\begin{equation}
    \bm{\mathcal{M}}_{ii} = \frac{1}{6\pi \eta a_i} \bm{I}
    \label{eq: rpy_diag}.
\end{equation}
Even though we restrict our problem to an unbounded geometry, it is possible to generalize the RPY tensor to other boundary conditions, such no-slip boundaries \cite{wajnryb_generalization_2013, swan_simulation_2007}, confined domains \cite{swan_particle_2010} and periodic domains \cite{mizerski_rotne-prager-yamakawa_2014, fiore_rapid_2017}. In the case of overlapping beads, some corrections need to be introduced to ensure that $\bm{\mathcal{M}}$ remains symmetric definite positive \cite{wajnryb_generalization_2013, zuk_rotneprageryamakawa_2014}, which is needed when Brownian motion is accounted for.

\section{Parameter sets}
\label{appendix:sets_of_parameters}
In the following table, we summarize the parameters used in our numerical simulations to compare with experiments. We recall that $\Delta s / \mathcal{C}$ is the obstacle cross-section shape resolution, where $\mathcal{C}$ is its contour length and $\Delta s $ the centerline distance between two consecutive beads. $\varepsilon^{-1} = L/2a$ is the fiber aspect-ratio, which is equivalent to its number of beads $N_F$. 

\begin{table}[ht]
    \begin{ruledtabular}
    \centering
    \begin{tabular}{ccc}
    Parameter                  & \textrm{Set I:  “Gliding event".}  & \textrm{Set II: “Trapping event".}\\
    \hline
          $\varepsilon^{-1}$  & 75    &  91      \\
          $a$[$\mu$m]    & 257    &  257      \\        
          $\Delta s / \mathcal{C}$        & 0.02    &   0.02     \\
          $\alpha $     & 0.61    &   0.61     \\
          $\omega$[cm]     & 1    &   1     \\
          $D_x/L$           & 0.25    &   0.03 \\
          $D_y/L$           & 1.75    &   1.45     \\
          $Be$          & 200    &   210     \\
          $\xi$         & 7.71    &   9.35     \\
          $\mathcal{K}$ & 0    &   0     \\
          $W$[N/cm]     & $13\cdot 10^{-6}$    &   $11\cdot 10^{-6}$      \\
          $\eta$[cP]       & 0.97    &   0.97     \\
    \end{tabular}
    \caption{Parameter sets used for comparison with experiments in Sec. \ref{effects_obstacle}}
    \label{tab : sets_of_parameters}
    \end{ruledtabular}
\end{table}

\newpage

\bibliography{main}

\end{document}